\documentclass[a4paper,12pt]{article}
\usepackage{graphicx}

\begin{document}

\centerline {\bf Quantum entanglement as information theoretic
resource}

\bigskip

\centerline{Miroljub Dugi\' c\footnote{E-mail address:
dugic@knez.uis.kg.ac.yu}}

\bigskip

\centerline{\it Department of Physics, Faculty of Science,}

\centerline{\it Kragujevac, Serbia and Montenegro}

\bigskip

{\bf Abstract:} We address the following criterion for
quantifying the quantum information resources: classically
simulable {\it vs.} classically non-simulable information
processing. This approach gives rise to existence of a deeper
level of quantum information processing--which we refer to as
"quantum communication channel". We particularly show, that
following the recipes of the standard theory of entanglement
measures does not necessarily give rise to un-locking the quantum
communication channel, which is naturally quantified by Bell
inequalities.

\bigskip


\vfill\eject

{\it Introduction}. - Quantum entanglement is recognized as a
resource for useful quantum information/computation (henceforth:
QIC) processing [1-6]. Consequently, quantifying entanglement is
among the central issues of QIC theory. The standard theory of
entanglement measures (cf., e.g., [1] for a review) offers the
recipes for improving entanglement (and/or purity) of a quantum
state (system) while bearing the following point as a background:
"the system is more entangled if it allows for better performance
of some task (impossible without entanglement)" [1]. However, it
is natural (for many reasons) to expect that certain additional
criteria/requirements concerning the state
preparation/manipulation might improve the performance of some
task. E.g. a more elaborate {\it information theoretic} analysis
might shed some new light in this concern; particularly, the
analysis devoted to the efficient (classically non-simulable)
tasks is not yet fully developed in the standard theory [1]. To
this end, it is useful to recall: distinguishing the classically
non-simulable yet quantum mechanically achievable processing is
probably the main motivation for developing QIC theory [7].

In this paper, we point out existence of the more fundamental
information processing level that we refer to as "entanglement as
information resource". Actually, we move the focus of the task of
quantifying entanglement to the following criterion/divide: {\it
classically simulable {\rm vs.} classically non-simulable}
information processing. This new approach directly addresses the
issue raised by Feynman [7] which is still un-resolved by the
standard theory of entanglement measures. With this, hopefully,
we sharpen the issue of necessity of entanglement for the
efficient (classically non-simulable) information processing.
Finally, recognizing "entanglement as information resource" as
the "quantum communication channel (QCC)", we may phrase our main
result as follows: manipulating entangled states according to the
recipes of the standard theory does not necessarily imply opening
(un-locking) {\it QCC}; in general, certain {\it additional
operations} are required for opening QCC, which as a quantum
information resource can be quantified by Bell-inequalities
(henceforth: BI) [8, 9].

\bigskip

{\it Quantum entanglement relativity.} - By definition, a
bipartite system is in entangled state if the state can not be
written in the {\it separable form} (with the straightforward
generalization to the mixed entangled states) [6, 10, 11 ]:

\begin{equation}
\vert \psi \rangle_1 \vert \chi \rangle_2,
\end{equation}

\noindent where $\vert \psi \rangle_1, \vert \chi \rangle_2$ are
the subsystems' states, and we omit the symbol of the tensor
product. Whilst this definition is as simple as clear, {\it
operationally} to justify entanglement is a bit subtle task.
E.g., if the $S$-factor of CHSH inequality [9] satisfies $S > 2$,
one may conclude that the system is in entangled state. However,
the result $S \le 2$ does {\it not} necessarily imply that the
state is of the separable form eq. (1). This is what we call {\it
"entanglement relativity"}. Namely, as it is well-known since the
pioneering paper of Bell [8], the trick is properly to choose the
observables to be measured {\it in respect} (i.e. {\it relative})
to the quantum state to be tested on entanglement.

By "entanglement relativity", we do not mean that quantum
entanglement is a relative concept--according to (1), the state is
either an entangled, or a separable state. What we have in mind is
the fact that, {\it operationally}, quantum entanglement need not
reveal itself. An example given in {\it Appendix A} clarifies our
notions.

Interestingly enough, this relativity of entanglement perfectly
fits with the Copenhagen interpretation of quantum mechanics
[12]: {\it depending on a physical situation} (here: of
measurement), a quantum state {\it either reveals}, {\it or} does
{\it not reveal} entanglement, which is here quantified by BI. To
this end, it is worth emphasizing: the operational estimation of
entanglement {\it we deal with} is only weakly linked to the
problem of deciding (by measurement) whether an unknown state is
entangled or not. The issue we have in mind is a bit more subtle.
Actually, and bearing in mind the {\it Principle of
Complementarity} [12], we want to emphasize that a quantum system
(in known or unknown entangled state) need not reveal
entanglement, very much like a quantum system need not reveal its
e.g. corpuscular nature (behaviour).

Extending this reasoning to the QIC protocols gives rise to some
interesting observations.

\bigskip

{\it Quantum teleportation: an analysis}. - In this section, we
strongly emphasize: quantifying the information resources cannot
rely solely on investigating the quantum system's states. Rather,
the information theoretic analysis of a process should reveal the
information resources. The following analysis of quantum
teleportation clarifies these notions.

Quantum teleportation is probably the most investigated QIC
protocol [3, 6, 13]. In the original paper [3], necessity of an
entangled state in the protocol has been pointed out. However, as
we point out in the sequel, there are still certain subtleties in
this regard.

As it is well-known, quantum teleportation can be described by the
stabilizer formalism [6, 14]. This formalism, in turn, can be
{\it efficiently simulated } on a {\it classical computer}--a
consequence of the profound Knill-Gottesman theorem [14]. That
is, while teleportation requires entangled systems [3], the
certain-states teleportation still can be classically {\it
simulated}--which is fundamental for our considerations.
Actually, as it is known since Bell [8], every classical
situation (here: the classical-computer simulation of the
stabilizer formalism) can be described by a {\it Local Hidden
Variables } ({\it LHV}) model, which, in turn, can be ascribed $S
\le 2$ [9]. At first sight, this may seem controversial: {\it
physically}, there is entanglement (described by the stabilizer
formalism) in the system implementing teleportation, yet the {\it
simulation} of teleportation (the classical-computer simulation of
the stabilizer formalism) is describable by $S \le 2$. However,
in analogy with "entanglement relativity", we may suppose that,
{\it operationally}, entanglement need not reveal itself. On the
other side, most quantum states can not be teleported by the use
of the stabilizer formalism [6, 14]. For such states, it is
expectable informaticly not to bear any {\it LHV} model thus
eventually giving rise to $S
> 2$.

Needless to say, physical processes are not identical with their
computer-simulation counterparts. It is therefore not for
surprise if the physical contents of a process do not reveal the
information theoretic contents (e.g. information resources) of
the process. Consequently, investigating the information resources
of a process should primarily rely on investigating the
information contents of a {\it simulation} of the process, as we
have essentially learned from Feynman [7]. In the above analysis:
quantum teleportation is a "physical process" employing entangled
states (systems), while the possible classical-computer
simulations of teleportation bear an {\it LHV} model. Certainly,
quantum teleportation as a {\it physical process} can not be
ascribed any {\it LHV} model, while certain {\it simulations} of
teleportation can be ascribed an {\it LHV} model.

These observations force us to conclude that {\it we should
distinguish} between "entanglement of a system (of quantum
hardware)" and the "entanglement as information resource". While
former refers to the physical systems (implementing the
information processing), the later refers to the information
contents of the processing on the deeper information theoretic
level that reveal the information resources. That is, operating
with an entangled {\it system} does not necessarily mean that, on
the more fundamental level of information processing,
entanglement as information {\it resource} has fully been
employed in the execution of the processing.

Now, the two tasks are in order. First, we should more closely
relate our findings to the standard theory of entanglement
measures. Second, we should offer some quantification of
"entanglement as information resource".

\bigskip

{\it Quantum information resources}. - Entangled quantum hardware
(entangled systems; entangled states) seems to be necessary for
the performance of (most of) the typical QIC tasks. On this level
of the information processing, entangled states appear as a
quantum information resource. This is exactly the issue of the
standard theory of entangled measures, which relies on the
definition of entanglement eq. (1). Physically, this resource may
be recognized as {\it quantum non-separability} [11]. To this
end, relying to the definition eq. (1) is as simple as clear a
criterion for non-separability, while quantifying
non-separability by BI raises some questions. Actually, violation
of BI (in quantum measurement) is linked with quantum nonlocality
[6, 8, 9], which, in turn, does not necessarily apply to all the
kinds of the entangled (non-separable) states--as it seems to be
the case with the bound entangled states.

However, our analysis of quantum teleportation points out
existence of the deeper level of information processing, which
employs entangled systems. In this regard, implicit to the
contents of the preceding section is the following
criterion/divide:

\medskip

{\it classically simulable {\rm vs.} classically non-simulable}
\hfill (2)

\medskip

\noindent {\it information processing}. Eq. (2) refers to the
possible {\it simulations} of QIC tasks and therefore (cf. the
preceding section) is suitable for deciding whether or not
"entanglement as information resource" has been employed in the
execution of the information processing, still bearing the
obvious {\it measure} (not yet in the mathematical sense):

$$S \le 2 \quad vs. \quad S > 2 \eqno(3)$$

\noindent in the order respective to (2); $S \le 2$ reveals un-use
while $S > 2$ reveals the possible use of "entanglement as
information resource".

In the anthropomorphic terms, "employing entanglement as
information resource" may be described as "opening (un-locking)
the quantum communication channel". Now, in analogy with
"entanglement relativity", we may read eqs. (2), (3) as follows:
the {\it choice} of {\it physical implementation} of a QIC
protocol gives, in principle, rise either to non-opening ($S \le
2$) or to un-locking ($S > 2$) the quantum communication channel
(QCC). In other words: quantum correlations (non-separability) in
a system do not {\it per se} constitute QCC--it is a matter of
physical situation whether or not this {\it virtue of entangled
systems} will be employed in the execution of the processing.
Based on eq. (2), this observation distinguishes QCC as the
possible basis for quantum protocols to beat the classical ones,
in analogy with the recently observed necessity and sufficiency of
violation of BI in the similar regard [15]. Needless to say,
non-opening of the quantum channel refers to the QIC tasks that
can be efficiently simulated on the "classical hardware".

Therefore, we may conclude that quantum entanglement bears (cf.
below) at least a double role as information resource: (i) quantum
non-separability of entangled systems, and (ii) the quantum
communication channel--the later being quantifiable by BI (the
r.h.s. of eq. (3)). While QCC apparently requires
non-separability, as we show in this paper, the later is not
sufficient for opening QCC.

Another information resource, quantum non-locality, is linked with
quantum non-separability--quantifying non-separability goes
through quantifying non-locality by (non)validity of BI [16], of
course except (probably) for the bound entangled states.
Therefore, QCC (as we introduce it in Section 4) is not
necessarily identical with the "quantum communication channel" of
the standard theory which is sometimes identified with quantum
non-locality. This is the reason we point out above the two,
mutually distinguishable resources--quantum non-separability and
QCC.

\bigskip

{\it Discussion}. - Following the recipes of the standard theory
of entanglement measures, one can improve entanglement in a
quantum system, yet without any guarantees about un-locking QCC
as information resource. The distinction between quantum
non-separability and QCC as the information resources is not quite
surprising yet. To this end, we have the following lessons in
mind. First, the classic lesson of the {\it Complementarity
Principle} gives rise to the expectation that, operationally,
entanglement need not reveal itself. Actually, in quantum
measurement, entanglement reveals itself in the specific quantum
measurement situations (cf. Section 2), while QCC reveals itself
through the classically non-simulable information processing (cf.
Sections 3, 4). Second, the true topic of QIC theory, as Feynman
has pointed it out [7] (cf. also [6]), is the performance of the
{\it classically non-simulable} tasks, which seems merely
un-tackled by the standard theory of entanglement measures.

Our approach to the issue of entanglement measures mainly refers
to QCC. It ultimately relies on the criterion (2), thus
presenting a general yet simple approach in quantifying
entanglement as information resource. By imposing the criterion
(2), we tackle the truly fundamental issue of QIC theory: 'whether
or not entanglement appears ultimate to efficiency of certain QIC
protocols/algorithms?'. Whilst the definite answer to this
question is a remote goal of the theory yet, classifying QIC
tasks in respect to (2) and (3) might help in setting the (e.g.,
empirically-based) recipe(s) for designing the efficient QIC
processing. E.g. the classical simulability of the superdense
coding, of the BB84 cryptographic protocol, as well as of the
certain-states teleportation clearly stems non-opening of QCC in
the course of the physical implementation of these protocols. It
is therefore easy to speculate about the future theory of
entanglement measures: the list of recipes from the standard
theory (referring to non-separability and/or probably to
non-locality) is {\it extended} by the recipes referring to the
opening of QCC--in an attempt to perform the classically both
inachievable and non-simulable information processing. Along this
line of reasoning, one may eventually clarify the relation between
quantum non-locality and QCC as the information resources: e.g.,
if it appears that entangled states non-violating BI can not be
used for the performance of the classically non-simulable tasks,
it might be interpreted in favor of identifying the two kinds of
the information resources--(ii) and (iii) in the above list of
resources.

Finally, making connection of our results with the standard
theory (cf. [1] and references therein) is not quite
straightforward a task. As yet, the relations of BI with the
standard entanglement measures (e.g., with the "concurrence" [17,
18]) are only weakly established [18] and only weakly understood
for the purposes of our considerations. Therefore, in this
respect, there is some research work yet to be done.

\bigskip

{\it Conclusion}. - We point out and discuss the following quantum
information resources linked with entanglement: (i)
non-separability (of quantum systems), (ii) quantum non-locality,
and (iii) quantum communication channel. The relation between the
first two resources is the true issue of the standard theory of
entanglement measures. The resource (iii) comes from the
addressing the following criterion for quantum information
resources: {\it classically simulable} vs. {\it classically
non-simulable} information processing--which is un-resolved by
the standard theory of entanglement measures. As we show,
manipulating entangled systems (non-separability, and probably
non-locality) does not necessarily mean that the quantum
communication channel as an information resource has been
employed in the execution of the information processing--which is
naturally quantified by Bell inequalities .

\bigskip

REFERENCES

\bigskip

[1] HORODECKI M., {\it Quantum Information and Computation} {\bf
1} (2001) 3

[2] BENNETT C. H. {\it et al}, {\it Phys. Rev. A} {\bf 54} (1997)
3814

[3] BENNETT C. H.  {\it et al}, {\it Phys. Rev. Lett.} {\bf 70}
(1993) 1895

[4] BENNETT C. H. and WIESNER S. J., {\it Phys. Rev. Lett.} {\bf
69} (1992) 2881

[5] SHOR P. W., In {\it Proceedings, 35th Annual Symposium on
Foundations of Computer Science}, IEEE Press, Los Alamitos, CA,
1994

[6] NIELSEN M. and CHUANG I., {\it "Quantum Computation and
Quantum Information"}, Cambridge University Press, Cambridge, UK,
2000

[7] FEYNMAN R. P., {\it Int. J. Theor. Phys.} {\bf 21} (1982) 467

[8] BELL J. S., {\it Physica} {\bf 1} (1964) 195

[9] CLAUSER J. F., HORNE M. A., SHIMONY A., HOLT R. A., {\it Phys.
Rev. Lett.} {\bf 49} (1969) 1804

[10] VON NEUMANN J., {\it "Mathematical Foundations of Quantum
Mechanics"}, Princeton University Press, Princeton, 1955

[11] D'ESPAGNAT B., {\it "Conceptual Foundations of Quantum
Mechanics"}, Reading, Mass., 1971

[12] BOHR N., {\it "Atomic Physics and Human Knowledge"}, Science
Edition, N.Y., 1961

[13] BOUWMEESTER D. et al, {\it Nature} {\bf 390} (1997) 575

[14] GOTTESMAN D., {\it Ph. D. thesis}, California Institute of
Technology, Pasadena, CA, 1997

[15] BRUKNER \v C. et al, {\it Phys. Rev. Lett.} {\bf 92} (2004)
127901

[16] WOOTERS W., {\it Quantum Information and Computation} {\bf 1}
(2001) 27

[17] BARRETT J. et al, e-print arxive quant-ph/0404097

[17] VERSTRAETE F. and WOLF M. M., {\it Phys. Rev. Lett.} {\bf 89}
(2002) 170401

[18] DUGI\' C M., {\it Eur. Phys. J. D} {\bf 29} (2004) 173

\bigskip

{\bf Appendix A}

\bigskip

Let us consider the following Bell states [6]:
$$\vert \psi^+ \rangle = 2^{-1/2} (\vert 0\rangle_1 \vert
1\rangle_2 + \vert 1\rangle_1 \vert 0\rangle_2), \quad \vert
\phi^+ \rangle = 2^{-1/2} (\vert 0\rangle_1 \vert 0\rangle_2 +
\vert 1\rangle_1 \vert 1\rangle_2) \eqno (A.1).
$$

{\it By definition}, these states are entangled pure states. Let
us now introduce the following set of the observables to be
measured on the composite (two-qubit) system:

$$\hat A_1 (\alpha) = 2^{-1} (\vert 0\rangle_1 + \exp(\imath \alpha) \vert 1\rangle_1)
(_1\langle 0\vert + \exp(- \imath \alpha) _1\langle 1\vert) -$$
$$2^{-1} (\vert 0\rangle_1 + \exp(\imath (\alpha + \pi)) \vert
1\rangle_1) (_1\langle 0\vert + \exp(- \imath (\alpha + \pi))
_1\langle 1\vert), \eqno (A.2)$$

$$\hat B_2 (\chi) = 2^{-1} (\vert 1\rangle_2 + \exp(\imath \chi) \vert 0\rangle_2)
(_2\langle 1\vert + \exp(- \imath \chi) _2\langle 0\vert) -$$
$$2^{-1} (\vert 1\rangle_2 + \exp(\imath (\chi + \pi)) \vert
0\rangle_2) (_2\langle 1\vert + \exp(- \imath (\chi + \pi))
_2\langle 0\vert), \eqno (A.3)$$

\noindent where the indices refer to the two qubits in the
composite system, and $\chi, \alpha \in [- \pi, \pi]$.

The standard $S$-factor of CHSH inequality [9] now reads:

$$S = E (\alpha_1, \chi_1) - E (\alpha_1, \chi_2) + E (\alpha_2, \chi_1)
+ E (\alpha_2, \chi_2), \eqno (A.4)$$

\noindent where $E(\alpha_i, \chi_j) = \langle \phi \vert \hat
A_1 (\alpha_i) \otimes \hat B_2 (\chi_j) \vert \phi \rangle, i,j =
1,2$.

In Fig. 1, we present the plot of the $S$-factor for $\vert \phi
\rangle = \vert \Psi^+ \rangle $, for which $E (\alpha, \chi) =
\cos (\alpha + \chi)$ [19], for the fixed values $\alpha_1 =
\pi/2$ and $\chi_1 = -\pi/4$.

\begin{figure}[!h]
\centering
\includegraphics[width=3.71in,height=2.32in]{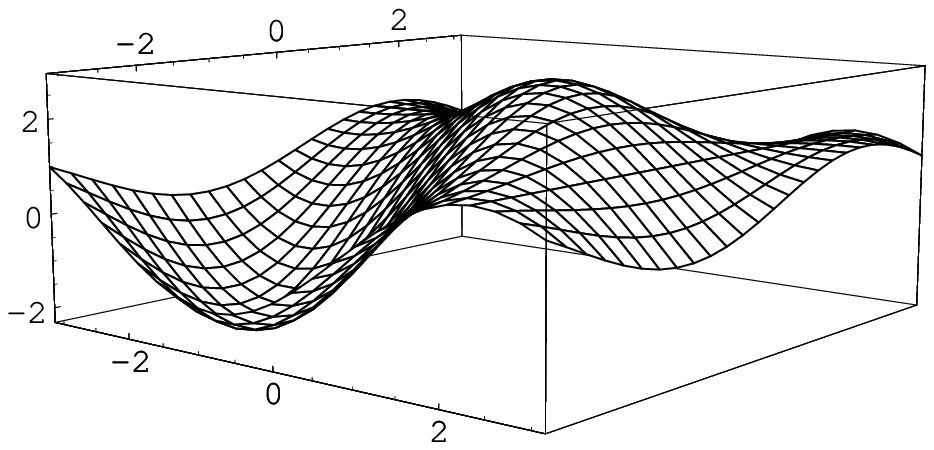}
\end{figure}

\centerline{Fig. 1}

\noindent The plot exhibits that the state $\vert \Psi^+
\rangle$, which is {\it by definition} entangled state, does {\it
not reveal entanglement} for the wrong choice of the observables
(here: of the angles $\alpha, \chi$). And this is exactly what we
mean by "entanglement relativity". The same conclusion applies to
other choices of $\alpha_1$ and $\chi_1$, as well as for the
state $\vert \phi^+ \rangle$.

\vfill\eject

FIGURE CAPTIONS

Fig. 1: The $S$-factor, eq. (A.4), for the state $\vert \Psi^+
\rangle$ and for fixed values $\alpha_1 = \pi/2, \chi_1 = -
\pi/4$. For the chosen (the maximal possible) value $S=2^{3/2}$,
the {\it plot returns} $\alpha_2 = 0$ and $\chi_2 = \pi/4$. For
$S = 0$, the plot returns e.g. $\alpha_2 = 0$, $\chi_2 = -
3\pi/4$. In general, the plot returns proportions of $\alpha_2,
\chi_2 $ for every fixed value of $S \in [2^{-3/2}, 2^{3/2}]$.

\end{document}